\documentstyle[aps,prl]{revtex}

\begin{document}

%%%%%%%%%%%%%%%%%%%%%%%%%%%%%%%%%%%%%%%%%%%%%%%%%%%%

\newcommand{\be}[1]{\begin{equation}\label{#1}}
\newcommand{\beq}{\begin{equation}}
\newcommand{\ee}{\end{equation}}
\newcommand{\beqn}[1]{\begin{eqnarray}\label{#1}}
\newcommand{\eeqn}{\end{eqnarray}}
\newcommand{\bd}{\begin{displaymath}}
\newcommand{\ed}{\end{displaymath}}
\newcommand{\mat}[4]{\left(\begin{array}{cc}{#1}&{#2}\\{#3}&{#4}\end{array}
\right)}
\newcommand{\matr}[9]{\left(\begin{array}{ccc}{#1}&{#2}&{#3}\\
{#4}&{#5}&{#6}\\{#7}&{#8}&{#9}\end{array}\right)}
\def\simlt{\mathrel{\lower2.5pt\vbox{\lineskip=0pt\baselineskip=0pt
           \hbox{$<$}\hbox{$\sim$}}}}
\def\simgt{\mathrel{\lower2.5pt\vbox{\lineskip=0pt\baselineskip=0pt
           \hbox{$>$}\hbox{$\sim$}}}}
\def\unity{{\hbox{1\kern-.9mm l}}}
\def\al{\alpha}
\def\ga{\gamma}
\def\Ga{\Gamma}
\def\om{\omega}
\def\OM{\Omega}
\def\la{\lambda}
\def\La{\Lambda}
\newcommand{\eps}{\varepsilon}
\def\ep{\epsilon}
\newcommand{\ov}{\overline}
\renewcommand{\to}{\rightarrow}
%%%%%%%%%%%% end my definitions %%%%%%%%%%%%%%%%%%%%%%%%%%%%%%%%%

\draft
\twocolumn[\hsize\textwidth\columnwidth\hsize\csname
@twocolumnfalse\endcsname
%\begin{frontmatter}
\vskip1pc

\title{ Unified picture of the particle and sparticle masses in SUSY GUT }
\author{Zurab Berezhiani} 
\address{INFN Sezione di Ferrara, I-44100 Ferrara, Italy, and \\ 
Institute 
of Physics, Georgian Academy of Sciences, 380077 Tbilisi, Georgia} 
%\date{23 August 1996}
\maketitle

\begin{abstract} 
The horizontal symmetry $U(3)_H$ can greatly help in solving 
the flavor problems in supersymmetric grand unification. 
We consider the $SU(5)\times U(3)_H$ model of ref. \cite{PLB85} 
and show that it can lead to the remarkable relations between 
the fermion mass matrices and the soft SUSY breaking terms 
at the GUT scale. 
%assuming the Planck scale universality of soft terms. 
The supersymmetric flavor-changing phenomena are naturally suppressed. 
The possible extensions to the $SO(10)\times U(3)_H$ model 
are also outlined.
\end{abstract}

%\end{frontmatter}

\vskip2pc]

%%%%%%%%%%%%%%%%%%%%%%%%%%%%%%%%%%%%%%%%%%%%%%%%%%%%%%%%%%%%

The flavor problem is one of the key problems in particle 
physics \cite{ICTP}. 
In the context of supersymmetric theories it has two aspects: 
fermion flavor and sfermion flavor. The first aspect concerns the 
quark and lepton mass spectrum and mixing pattern. 
In the minimal supersymmetric standard model (MSSM) the fermion 
sector $f$ consists of three  families 
$l_i\!=\!(\nu,e)_i$, $e^c_i$, $q_i\!=\!(u,d)_i$, $u^c_i,d^c_i$
($i=1,2,3$). Their masses emerge from the Yukawa couplings 
to the Higgs doublets $H_{1,2}$: 
\be{Yuk} 
W_{\rm Y}= l \hat{\la}_e e^c H_1 + q \hat{\la}_d d^c H_1 
+ q \hat{\la}_u u^c H_2 , 
\ee 
where the $3\times 3$ matrices  $\hat{\la}_{e,d,u}$ of the coupling 
constants remain arbitrary. However, 
supersymmetric grand unified theories (SUSY GUT), in particular, 
the models \cite{so10,su6} based on $SO(10)$ or $SU(6)$ symmetries 
can provide predictive textures for the Yukawa couplings.

The second aspect, a specific of SUSY, questions the mass and mixing 
spectrum of (yet undiscovered) superpartners, 
or in other words the pattern of the soft SUSY breaking (SSB) terms 
(the tilde labels sfermions): 
\beqn{SSB} 
& L_A = \tilde{l} \hat{A}_e \tilde{e}^c H_1 
+ \tilde{q} \hat{A}_d \tilde{d}^c H_1 
+ \tilde{q} \hat{A}_u \tilde{u}^c H_2,  \nonumber \\ 
& L_m = \sum_f \tilde{f}^\dagger \hat{m}^2_{\tilde{f}} \tilde{f} 
~~~~~ (f=l,e^c,q,u^c,d^c) .
\eeqn 
Here $\hat{A}_{e,d,u}$ and $\hat{m}^2_{\tilde f}$ are the arbitrary 
$3\times 3$ matrices, not necessarily aligned to the Yukawa 
constants in (\ref{Yuk}).   
%with the $\sim m_S$ and $\sim m_S^2$ elements respectively, 
%where $m_S$ of few hundred GeV is a SUSY breaking scale.  
Hence, the fermion-sfermion couplings to neutral gauginos 
in general are not diagonal in the basis of the mass eigenstates.  
This induces the flavor changing (FC) processes 
like $K^0-\bar K^0$ transition, $\mu\to e\ga$ decay,   
%CP-violating parameter $\eps_K$ in $K^0-\bar K^0$ system, etc., 
with the rates much exceeding the experimental ones unless the 
sfermion masses are about 100 TeV or larger. In this case,  
however, the advantages of supersymmetry in 
stabilizing the Higgs boson mass would be lost.  

In MSSM natural suppression of the SUSY FC phenomena can be 
achieved by assuming the SSB terms universality at the Planck
scale \cite{BFS}. 
However, in the context of SUSY GUTs this idea becomes insufficient.  
The FC effects of physics above the GUT scale $M_X$ can strongly 
violate the soft-terms universality at lower scales \cite{HKR}.   
This has most dramatic impact on the predictive frameworks 
like \cite{so10,su6}. 
These theories, in addition to the standard chiral set of the 
`would be light' fermion superfields  $f=q,u^c,d^c,l,e^c$ and 
standard Higgses $H_{1,2}$, contain also a vector-like set of heavy  
fermions $F+F_c$ 
%in representations similar to that of $f$'s   
and various GUT Higgses. Typically all these have masses (VEVs) 
$\geq M_X$ and large couplings (with constants $\sim 1$).  
Actual identification of the fermion flavor 
occurs after integrating out the heavy sector \cite{FN}, when 
theory reduces to the MSSM, and the light physical states $f'$ 
generically emerge as linear combinations 
of the original $f$ and $F$ states. 
This feature provides an appealing explanation to the fermion flavor 
structure, since the small Yukawa constants in the MSSM can be 
understood as the ratios of different physical scales in 
theory \cite{FN,so10,su6}. 
However, the same feature creates severe problems in the sfermion sector. 
Even if the SSB terms were universal at $M_{Pl}$, 
their universality at energies below $M_X$ 
can be strongly violated due to effects of the renormalization 
group (RG) running from $M_{Pl}$ down to $M_X$ \cite{HKR} 
and the heavy sector decoupling \cite{DP}. 
This would cause dramatic FC phenomena 
which pose a serious challenge to a whole concept of 
the supersymmetric grand unification.

A natural way to approach both flavor problems is to 
introduce the horizontal (inter-family) symmetry.  
Several possibilities based on the flavor symmetries 
$U(1)$ or $U(2)$ have been considered in the literature \cite{NS}.  
However, the chiral $SU(3)_H$ symmetry \cite{su3H} unifying all 
families in horizontal triplets seems to be the most attractive candidate. 
%for describing the family replication \cite{su3H}.  
In this letter we revisit the $SU(5)\times SU(3)_H$ model 
suggested earlier in ref. \cite{PLB85}. We show that it can provide 
unified picture of the particle/sparticle masses 
and naturally solve the supersymmetric FC problem. 

The model contains $f$ fermions in the representations 
\be{f}
f_{\bar5}=(d^c\!, l)_i \sim (\bar5,3), ~~~~ 
f_{10}=(u^c\!, q, e^c)_i \sim (10,3)  
\ee 
($i=1,2,3$ is the $SU(3)_H$ index), while the standard Higgses 
$H_{1,2}$ are placed respectively in the superfields 
$\bar H\sim (\bar5,1)$ and $H\sim (5,1)$. 
Since the fermion mass terms transform as $3\times 3=6+\bar3$, they 
can be induced only via the higher order operators 
\be{HOPs}
G_n\frac{\chi_n}{M} f_{\bar5}f_{10} \bar H + 
G'_n\frac{\chi_n}{M} f_{10} f_{10} H , 
\ee 
where $M\gg M_W$ is some cutoff scale (hereafter to be referred to as a 
flavor scale) and $\chi_n$ ($n=1,2,$...) denote the 
Higgs superfields in two-index representations of $SU(3)_H$:   
(anti)sextets $\chi^{\{ij\}}$ and triplets 
$\chi^{[ij]}\sim \eps^{ijk}\chi_k$, 
with VEVs $\langle \chi_n \rangle < M$. 
%(With respect to $SU(5)$ these can transform as $1, 24$ or 75). 
Notice, that the actual global chiral symmetry of terms (\ref{HOPs}) 
is $U(3)_H=SU(3)_H\times U(1)_H$, where $U(1)_H$ related to 
the phase transformations $f_{\bar5},f_{10}\to e^{i\om}f_{\bar5},f_{10}$;  
$\chi_n\to e^{-2i\om}\chi_n$, can serve as 
Peccei-Quinn symmetry unless it is explicitly 
broken in the Higgs potential of $\chi_n$ \cite{PLB85}. 

This picture suggests that 
the observed mass hierarchy may emerge from the hierarchy 
${\cal C}\gg {\cal B} \gg {\cal A}$ of the chiral $U(3)_H$ symmetry 
breaking scales:
\be{chain}
U(3)_H \stackrel{{\cal C}}{\rightarrow}U(2)_H \stackrel{{\cal B}}
{\rightarrow} U(1)_H \stackrel{{\cal A}}{\rightarrow} I
\ee 
rather than due to the {\em ad hoc} choice of small Yukawa constants. 
In addition, a specific form of the VEVs $\langle \chi_n \rangle$
can lead to the predictive textures for fermion masses. 
For example, one can choose $\chi_n$ as a sextet $\chi_3$ 
having a VEV ${\cal C}$ towards (3,3) component  
and two triplets $\chi_{2,1}$ with VEVs ${\cal B}$ and ${\cal A}$ 
pointing respectively to the 1$^{st}$ and 3$^{rd}$ directions.   
Then the whole VEV matrix has a Fritzsch texture \cite{Fr}  
to be transferred also to fermion mass terms: 
\be{H-VEVs} 
\hat{V}_H=\sum_n \langle \chi_n \rangle = 
\matr{0}{{\cal A}}{0} {-{\cal A}}{0}{{\cal B}} {0}{-{\cal B}}{{\cal C}} . 
\ee
In the following this pattern 
will be used as a simple qualitative reference model \cite{susy96}. 
In particular, it implies the following hierarchy of 
the $U(3)_H$ breaking scales: 
\be{ABC} 
{\cal A}:{\cal B}:{\cal C}\sim \sqrt{m_1m_2} : \sqrt{m_2m_3} : m_3, 
~~~~ {\cal C}\sim M ,  
\ee
where $m_{1,2,3}$ are typical masses of three fermion families. 

The unbroken $U(3)_H$ would imply the mass degeneracy between 
the sfermion families. However, once nonrenormalizable terms 
(\ref{HOPs}) are included in the superpotential, 
there is no symmetry reason to exclude the analogous operators 
in the SUSY breaking sector. Namely, there can be the following 
terms with arbitrary coefficients: 
\beqn{SSB-hops}
& A_n\frac{\chi_n}{M} \tilde{f}_{\bar5} \tilde{f}_{10} \bar H 
+ A'_n\frac{\chi_n}{M} \tilde{f}_{10} \tilde{f}_{10} H , \nonumber \\ 
& m_5^2 \tilde{f}_{\bar5}^\dagger
\big (1 + \gamma_{kn}\frac{\chi_k^+\chi_n}{M^2}\big) \tilde{f}_{\bar5}  
+ m^2 \tilde{f}_{10}^\dagger 
\big(1+ \gamma'_{kn}\frac{\chi_k^+\chi_n}{M^2} \big) \tilde{f}_{10}. 
\eeqn 
The SSB terms (\ref{SSB}) which emerge after substituting the VEVs 
$\langle \chi_n \rangle$ are non-universal in families. 
In view of estimates (\ref{ABC}) the mass splittings and mixings 
between sfermions are significant, so that the contributions 
e.g. to the CP-violating parameter $\eps_K$ in $K^0-\bar K^0$ 
system will be too large unless the sparticle mass scale is of few TeV. 
This situation, generic also for the models \cite{NS}, 
however implies the certain amount of fine tuning for 
stabilizing the Higgs boson mass. Therefore, 
more convincing approach would be desirable. 

It was suggested in \cite{PLB85} that operators (\ref{HOPs}) emerge 
entirely from the renormalizable theory, after integrating out some 
heavy (with mass $\sim M$) $F$-states. Let us introduce the 
vector-like fermions in representations 
\be{F}
F\! =\! (U^c\!,Q,E^c)^i\! \sim (10,\bar 3), ~~ 
F_c\! =(U,Q^c\!,E)_i \! \sim (\ov{10},3)   \nonumber \\ 
\ee
and consider the most general Yukawa terms: 
\be{Rterms}
{\cal W}_Y=
gf_{\bar5}F\bar H +g'f_{10}FH +b\Sigma FF_c 
+ a_n\chi_n F_cf_{10} ,
\ee
where the `horizontal' Higgses $\chi_n\sim (R,r)$ can be in all 
possible representations of $SU(5)\times SU(3)_H$, 
with $R\subset 10\times \ov{10}=1+24+75$ and 
$r\subset \bar3 \times \bar3=3+\bar6$. 
For simplicity we take $\Sigma$ as a singlet, with the VEV 
$\langle \Sigma \rangle = M\sim M_X$. 
The general pattern of the SSB terms is 
\beqn{tri-F}
& {\cal L}_A = 
A\tilde{f}_{\bar5} \tilde{F} \bar H + A'\tilde{f}_{10}\tilde{F} H 
+ B\Sigma\tilde{F}\tilde{F}_c + Aa'_n\chi_n \tilde{F}_c\tilde{f}_{10}    
\nonumber \\ 
& {\cal L}_m = m^2_5 \tilde{f}_{\bar5}^\dagger \tilde{f}_{\bar5} + 
m^2 \tilde{f}_{10}^\dagger \tilde{f}_{10} + 
\mu^2 \tilde{F}^\dagger \tilde{F} + 
\mu_{c}^2 \tilde{F}_c^\dagger \tilde{F_c} 
\eeqn 
where the magnitudes of the dimensional parameters 
%$A,A',...$ and $m,m_5,...$ 
can vary from hundred GeVs to TeV.

After substituting the VEVs $\langle \chi_n \rangle$ 
couplings (\ref{Rterms}) reduce to the following mass matrices:   
\beqn{Ml}
&  \begin{array}{cc}
 & {\begin{array}{cc} \,e^c & \,\,\,\;E^c \end{array}}\\ \vspace{2mm}
\begin{array}{c}
l \\ E \end{array}\!\!\!\!\!&{\left(\begin{array}{cc}
0 & \hat{g}H_1 \\ \hat{M}_e  & \hat{M} \end{array}\right)} 
\end{array} \! , ~~~~~~
\begin{array}{cc}
 & {\begin{array}{cc} \,d^c & \,\,\,\;Q^c \end{array}}\\ \vspace{2mm}
\begin{array}{c}
q \\ Q \end{array}\!\!\!\!\!&{\left(\begin{array}{cc}
0 &  \hat{M}_q^T \\ \hat{g}H_1 & \hat{M} \end{array}\right)} 
\end{array} \! ,    \nonumber \\ 
&  \begin{array}{ccc}
 & {\begin{array}{ccc} \,u^c & \,\,\,\;Q^c & \,\,\;U^c 
\end{array}}\\ \vspace{2mm}
\begin{array}{c}
q \\ Q \\ U   \end{array}\!\!\!\!\!&{\left(\begin{array}{ccc}
0 & \hat{M}_q^T & \hat{g}'H_2 \\ 
\hat{g}'H_2 & \hat{M} & 0 \\ 
\hat{M}_u  & 0 & \hat{M}  \end{array}\right)} 
\end{array} .
\eeqn 
where each entry is $3\times 3$ matrix in itself: 
$\hat{g}=g\unity$, $\hat{g}'=g'\unity$ and $\hat{M}=M\unity$  
are the family blind matrices, and all information on the 
fermion flavor pattern is contained in the off-diagonal blocks 
$\hat{M}_{f}= \sum_n a_n \langle \chi_n \rangle_{f}$. 
%reflect the structure of the VEVs $\langle \chi_n \rangle$: 
Here the subscript $f=e,q,u$ implies that the fields $\chi_n$, 
depending on their $SU(5)$ content, induce different Clebsches 
in corresponding elements of  the matrices $\hat{M}_{e,q,u}$. 

After integrating out the heavy sector theory reduces to 
the MSSM with the light fermion superfields $f'$ 
%($f'=q',l',u_c',d_c',e_c'$) 
being the linear combinations of the initial $f$ and $F$ states. 
Notice, that $d^c$ and $l$ (components of $f_{\bar5}$) 
do not mix to the heavy states: $l'=l$, $d'^c=d^c$,  
while the members of $f_{10}$ ($u^c,q,e^c$) mix to the 
corresponding states ($U^c,Q,E^c$) in $F$. In the  `seesaw' limit 
$\hat{M}_{q,u,e}\ll M$ 
%the light states are given as $q'\simeq q -\frac{1}{M}\hat{M}_qQ$, etc. 
%(in matrix notations).
this yields the following form of the MSSM Yukawa matrices (\ref{Yuk}): 
\be{Yukawas} 
\hat{\la}_e = g\hat{\kappa}_e, ~~~ 
\hat{\la}_d = g\hat{\kappa}_q^T, ~~ 
\hat{\la}_u = g'(\hat{\kappa}_q^T + \hat{\kappa}_u) .
\ee
where $\hat{\kappa}_f= \hat{M}^{-1}\hat{M}_f$, $f=e,q,u$.  
For example, one can fix the sextet $\chi_3$ inducing the largest VEV 
(${\cal C}$) in the $U(3)_H$ symmetry breaking chain (\ref{chain}) 
as $SU(5)$ singlet: $\chi_3\sim (1,\bar6)$. 
Then the (3,3) entries of $\hat{M}_{q,u,e}$ are equal and thus we 
maintain the $b-\tau$ Yukawa unification at the GUT scale: 
$\la_b=\la_\tau$,  while $\la_t/\la_b=2g'/g$ \cite{Rattazzi}.  
The fields $\chi_n$ with smaller VEVs 
should include the ones in the mixed representations  
like $\chi_2\sim (24,3)$, in order to produce the 
nontrivial $SU(5)$ Clebsches distinguishing the quark and lepton 
mass entries for the light families \cite{HOP}. 
In this paper we do not assume any particular texture for the 
VEVs $\langle \chi_n \rangle$: 
$\hat{M}_{q,u,e}$ are general matrices with a hierarchy of elements 
that is typical for the fermion mass matrices.  
However, by imposing the certain simple textures  
on $\langle \chi_n \rangle$,  e.g. (\ref{H-VEVs}) or alike, 
one could produce fermion mass ansatzes with a predictive 
power as strong as in the $SO(10)$ based models \cite{so10}. 
An example of such a predictive model 
was presented in ref. \cite{susy96}. 

Substituting the VEVs $\langle \chi_n \rangle$ in the full scalar 
Lagrangian, 
we obtain the total mass matrix e.g. for the slepton states 
$(\tilde{l}^*, \tilde{E}^*, \tilde{e}^c, \tilde{E}^c)$:  
\be{slept}
\left(  \begin{array}{cccc}
\hat{m}^2_5 & 0 & 0 & \hat{A} H_1  \\  
0 & \hat{\mu}_{c}^2 \!+\! \hat{M}^2 \!+\! \hat{M}_e\hat{M}_e^\dagger & 
 A\widetilde{M}_e & B \hat{M}  \\ 
0 & A\widetilde{M}_e^\dagger & 
\hat{m}^2 \!+\! \hat{M}_e^\dagger \hat{M}_e & M\hat{M}_e \\ 
\hat{A}H_1^* & B\hat{M} & M \hat{M}_e^\dagger & 
\hat{\mu}^2 \!+\! \hat{M}^2 
\end{array}\right)  
\ee
and analogously for squarks, where $\hat{m}^2 = m^2\unity$, 
$\hat{A} = A\unity$, etc. are family blind entries and 
$\widetilde{M}_{f}= \sum_n a'_n \langle \chi_n \rangle_f$, 
$f=e,q,u$,   
have approximately the same pattern as $\hat{M}_{e,q,u}$.  
%(hereafter we omit $'$ for the light (MSSM) states) 
Integrating out the heavy fields in matrices (\ref{slept}), 
one can obtain the SSB terms (\ref{SSB}) which are left 
for the light $f'$ states in the MSSM. 
We find that all $A$-terms are aligned to the corresponding Yukawa 
matrices:  
\be{A} 
\hat{A}_e = \frac{A}{g}\hat{\la}_e , ~~~ 
\hat{A}_d = \frac{A}{g} \hat{\la}_d , ~~~ 
\hat{A}_u = \frac{A'}{g'}\hat{\la}_u ,  
\ee 
The fragments of $f_{\bar5}$, $d^c$ and $l$, do not mix to the 
heavy states and their soft masses remain family blind:  
\be{ldc}
\hat{m}^2_{\tilde{l}} = m^2_5\unity , ~~~~~~~ 
\hat{m}^2_{\tilde{d}^c} = m^2_5\unity , 
\ee
while the other soft mass matrices are the following: 
\beqn{softq}
& \hat{m}^2_{\tilde{e}^c}=m^2\big(\unity 
- \delta\hat{\kappa}_e^\dagger \hat{\kappa}_e 
- \delta_A\hat{\rho}_e^\dagger\hat{\rho}_e \big) , 
 \nonumber \\
& \hat{m}^{2}_{\tilde{q}} = m^2\big(\unity -  
\delta \hat{\kappa}_q^\dagger \hat{\kappa}_q 
- \delta_A \hat{\rho}_q^\dagger\hat{\rho}_q \big) , 
 \nonumber \\ 
& \hat{m}^2_{\tilde{u}^c} = m^2\big(\unity - 
\delta\hat{\kappa}_u^\dagger \hat{\kappa}_u 
- \delta_A \hat{\rho}_u^\dagger\hat{\rho}_u \big) ,
\eeqn
where $\delta= (\mu^2 - m^2)/m^{2}$, $\delta_A = A^2/m^2$ 
are generally $\sim 1$. The matrices 
$\hat{\rho}_f=\frac{1}{M}\widetilde{M}_f -\frac{B}{A}\hat{\kappa}_f$  
are approximately aligned to $\hat{\kappa}_f$, but are not exactly 
proportional unless $a'_n\propto a_n$. 
The terms $\sim \delta$ arise due to difference of the soft masses 
$m$ and $\mu$, from the same rotation of superfields in $f_{10}$ 
and $F$ which defines 
the physical light ($f'_{10}$) and heavy ($F'$) fermions. 
The terms $\sim \delta_A$ emerge due to the additional small mixing 
($\sim A/M$) between the states $\tilde{f}'_{10}$ and $\tilde{F}_c^*$.  
In fact,  these are contributions to the soft masses 
of the physical states which emerge from the $A$-terms 
(\ref{tri-F}) in decoupling the heavy sector. 

Thus, in our model the boundary conditions for the SSB terms at the 
flavor scale are different from the {\em universal} ones usually  
adopted in the MSSM. In order to deduce the SSB terms pattern at 
the electroweak scale,  the expressions (\ref{A}), (\ref{ldc}) and 
(\ref{softq}) should be corrected by the RG running below the scale $M$. 
Clearly, all FC effects will remain under control, once they are 
not big at the scale $M$. 
Namely, eqs. (\ref{A}) and (\ref{ldc}) imply that there are 
no flavor changing in the left sleptons and right down squarks, 
as well as in L-R mixing. This leads to  immediate suppression 
of the supersymmetric contributions which are generally dominant 
to the $K^0-\bar{K}^0$ transition and $\mu \to e\gamma$ decay.  
%(gluino mediated boxes involving $\tilde{d}^c$ states) 
%(loops involving $\tilde{l}$ states) are vanishing.  
Nevertheless, the contributions to $\eps_K$ and electric dipole 
moments can be still too big for the sfermion masses up to 
$1$ TeV unless the CP-violating phases are small. 
The latter case could occur e.g. if the theory possesses 
CP-invariance which is spontaneously broken by the `horizontal' 
VEVs $\langle \chi_n \rangle$. 

It is interesting to consider a situation when $\hat{\rho}_f$ 
are vanishing. This would occur if the SSB terms 
were universal at the Planck scale, 
i.e. ${\cal L}_A \propto {\cal W}_Y$ \cite{BFS}. 
As it was shown in \cite{DP}, the $A$-terms RG running below $M_{Pl}$ 
is not relevant at the decoupling of the heavy sector, which 
fact actually is an implication of the F-terms no-renormalization. 
Then, taking into account eqs. (\ref{Yukawas}), we obtain \cite{prop}: 
\beqn{softm}
& \hat{m}^2_{\tilde{e}^c}=m^2\big(\unity 
- \frac{\delta}{g^2} \hat{\la}_e^\dagger \hat{\la}_e \big) , ~~~
 \hat{m}^{2}_{\tilde{q}} = m^2\big(\unity -  
\frac{\delta}{g^2} \hat{\la}_d \hat{\la}_d^\dagger \big)^T , 
\nonumber \\
& \hat{m}^2_{\tilde{u}^c} = m^2\big[\unity - 
\frac{\delta}{g^2} 
\big(\hat{\la}_d^\dagger - \frac{g}{g'}\hat{\la}_u^\dagger \big)   
\big(\hat{\la}_d - \frac{g}{g'}\hat{\la}_u \big)  \big]
\eeqn
(recall also that $g/g'=2\la_b/\la_t$ if $\chi_3 \sim (1,\bar6)$). 
Thus, in this case whole pattern of the SSB couplings (\ref{SSB}) 
at the scale $M$ is completely fixed 
solely in terms of the Yukawa matrices (\ref{Yuk}) 
irrespective of the concrete form of the latter. 
(Note, for $\delta >0$ the sfermion masses (\ref{softm}) exhibite  
the inverse hierarchy). 
One can rotate fermion states to a basis in which 
$\hat{\la}_{e}$ and $\hat{\la}_{d}$ are diagonal: 
$\hat{\la}_{e,d}=\hat{\la}_{e,d}^D$  
%$\hat{\la}_{e}=\hat{\la}_{e}^D={\rm Diag}(\la_e,\la_\mu,\la_\tau)$, 
%$\hat{\la}_{d}=\hat{\la}_{d}^D={\rm Diag}(\la_d,\la_s,\la_b)$  
and $\hat{\la}_{u}=V^\dagger \hat{\la}_{u}^D V_c$, where 
%$\hat{\la}_{u}^D={\rm Diag}(\la_u,\la_c,\la_t)$ and 
$V$ is the CKM matrix. 
Hence,  all {\em sflavor} parameters 
(15 sfermion masses and 7 matrices of the $f-\tilde{f}$ mixing) 
are predicted in terms of 13 standard parameters 
(fermion masses and CKM matrix) plus 5 new parameters: 
masses $m,m_5$,  dimensional constants $A,A'$ and dimensionless 
$\delta/g^2$ defining the relative mass splittings between sfermion 
families.  For example, we have mass relations 
\be{smasses} 
m^2_{\tilde{e}}=m^2_{\tilde{\mu}}=m^2_{\tilde{\tau}}, ~~~~
\frac{ m^2_{\tilde{\mu}^c} - m^2_{\tilde{e}^c} }
{m^2_{\tilde{\tau}^c}-m^2_{\tilde{e}^c} }= 
\frac{m^2_\mu}{m^2_\tau} 
\ee
for sleptons and the similar formulas for squarks. 

Now the soft mass terms of the $\tilde{e}^c$ and $\tilde{d}\subset q$ 
states as well as the trilinear couplings $\hat{A}_e$ and $\hat{A}_d$ 
are aligned respectively to $\hat{\la}_e$ and $\hat{\la}_d$, 
while the states $\tilde{l}$ and $\tilde{d}^c$ are degenerate between 
families.  Therefore, 
there are no new supersymmetric contributions to the 
%mediated by exchanges of the down squarks or sleptons: 
$K^0-\bar{K}^0$ transition, $b\to s\gamma$ and $\mu\to e\ga$ decays, 
etc. (besides the standard ones induced due to the RG running 
from the scale $M$ down to the electroweak scale).  
The new FC effects can emerge only in the upper sector.  
Since $\hat{m}^2_{\tilde{q}}$ is aligned to $\hat{\la}_d$, 
the mixing angles in the $u$-$\tilde{u}$-gluino couplings 
coincide with the CKM angles. The magnitudes of the $u^c$-$\tilde{u}^c$ 
mixing angles also have the size of the CKM angles.  
Therefore, predictions for $D^0-\bar{D}^0$ transition, etc. 
are consistent with the current experimental limits.  

In principle, the flavor scale $M$ can be different from the GUT 
scale $M_X\simeq 10^{16}$ GeV. However, it can be constrained by 
the following arguments. 
The contributions of the extra superfields (\ref{F}) 
would ``explode'' (violate perturbativity bound) 
in the $SU(5)$ gauge constant running before $M_{Pl}$ unless 
$M>10^{12}$ GeV. As we have noted, 
the field $\chi_{2}$ should be in the mixed representation 
(24,3) or alike, and thus ${\cal B} \leq M_X$.  
Interestingly, 
the scale ${\cal B}\sim 10^{16}$ GeV is also favored if one attempts 
to explain the large scale structure of the universe via the global 
textures related to the $SU(3)_H$ symmetry breaking \cite{Turok}.
The lower limit 
%on the intermediate $U(2)_H$ breaking scale 
${\cal B}> 10^{10}$ GeV emerges from the experimental bound 
on the decay $K^+ \to \pi^+$ + {\em familon} \cite{Kim}.  
% these bounds imply that $M =10^{11}-10^{17}$ ${\rm GeV}$. 
In addition, if $U(1)_H$ is not explicitly broken in the Higgs 
sector, then it plays a role of the Peccei-Quinn symmetry  
and astrophysical constraints on the axion imply that 
${\cal A} = 10^{10}-10^{12}$ GeV \cite{Kim}. 
Taking into account the estimates (\ref{ABC}), all these bounds 
translate into $M= 10^{13}-10^{16}$ GeV. 
%$10^{13}~{\rm GeV} < M < 10^{16}~{\rm GeV}$.
As a matter of fact, 
these constraints ensure that possible contributions from 
%the stability of our picture against possible 
the truly nonrenormalizable Planck scale 
operators like (\ref{HOPs}) and (\ref{SSB-hops}) 
would be negligible both for the fermion masses and the SSB terms. 

%%%%%%%%%%%%%%%%%%%%%%%%%%%%%%%%%%%% SO(10) %%%%%%%%%%%%%%

One can extend the symmetry group to $SO(10)\times U(3)_H$, 
In this case the $f$-fermions (\ref{f}) are included in
$f \sim (16,3)$ while the Higgses $H, \bar H$ are combined 
in $h \sim (10,1)$. 
As for the $F$-fermion set (\ref{F}), it should be enlarged 
to $F\sim (16,\bar3)$ and $F_c\sim (\ov{16},3)$, 
which now include also the $SU(5)$ 5-plets 
$F' \! =\! (D^c,L)^i \! \sim (\bar5,\bar 3)$ and   
$F'_c \!=\!(D,L^c)_i \! \sim (5,3)$. (Certainly,  
the latter could be introduced directly in the 
$SU(5)\times U(3)_H$ frames \cite{PLB85}, but such a theory 
would not be very predictive.) 

One can consider models with the most general pattern of 
the renormalizable Yukawa terms: 
\be{W-so10}
{\cal W}_Y^{SO(10)}=gfFh + b_k\Sigma_k FF_c + a_n \chi_n F_c f,  
%\nonumber \\ 
%& {\cal L}_A^{SO(10)}= A\tilde{f}\tilde{F}h 
%+ B\Sigma \tilde{F}\tilde{F}_c 
%+ A_n \chi_n\tilde{F}_c \tilde{f} ,  \nonumber \\ 
%& {\cal L}_m^{SO(10)}= m^2 \tilde{f}^\dagger \tilde{f} 
%+ \mu^2 \tilde{F}^\dagger \tilde{F} 
%+ \mu^2_{c} \tilde{F}_c^\dagger \tilde{F}_c ,   
\ee   
where $\chi_n\sim (R,\bar6+3)$ and $\Sigma_k \sim (R,1+8)$ stand for all 
possible Higgses with $R=1,45$ or 210. 
%However, for simplicity 
%we take only $\Sigma \sim (1,1)$ with $\langle \Sigma \rangle = M$.
%
After substituting the scalar VEVs, 
mass matrices of the $f$ and $F$ fermions all obtain the form 
similar to the upper quark matrix in (\ref{Ml}), involving now 
 nontrivial entries $\hat{M}_f$ and  $\hat{M}_F$  which are  
defined by the VEV structures $\langle \chi_n \rangle$ and 
$\langle \Sigma_k \rangle$. 
After integrating out the heavy sector, all the 
Yukawa and SSB couplings will be fixed in terms of 5 matrices 
$\hat{\kappa}_f= \hat{M}_F^{-1}\hat{M}_f$ ($f=q,u,d,e,l$):
\beqn{so10}
& \hat{\la}_e = g(\hat{\kappa}_l^T \!+\! \hat{\kappa}_e) , ~~~~~~ 
%\hat{\la}_d = g(\hat{\kappa}_q^T \!+\! \hat{\kappa}_d), ~~~ 
\hat{\la}_{d,u} = g(\hat{\kappa}_q^T \!+\! \hat{\kappa}_{d,u}), 
\nonumber \\ 
& \hat{A}_{e,d,u}=\frac{A}{g}\hat{\la}_{e,d,u} , ~~~~ 
\hat{m}^2_{\tilde{f}}=m^2 (\unity - 
\delta\hat{\kappa}_f^+\hat{\kappa}_f) 
\eeqn  
(Note, the last relation is obtained assuming the Planck scale 
universality of the SSB terms). By imposing certain zero textures 
like (\ref{H-VEVs}) on the VEVs $\langle \chi_n \rangle$, one can obtain 
the matrices $\hat{\kappa}_f$ with reduced number of parameters, 
where the corresponding entries 
for different states differ only by the $SO(10)$ Clebsch factors. 
In this way one can produce the highly predictive ansatzes 
for fermion masses,  in the spirit of the models \cite{so10}. 
However, in contrast to the latter schemes, the FC processes can be 
naturally controlled in these models, and what is 
more interesting, they can be tested also in the sfermion 
sector which, according to (\ref{so10}), involves only 
two new dimensional parameters $A/g$ and $\delta/g^2$. Note, that 
non of the soft mass matrices in (\ref{so10}) is aligned precisely 
to the Yukawa terms, and the fermion-sfermion mixing angles 
generally are expected to be of the order of the CKM angles. 
Therefore, these models can have testable predictions for 
the supersymmetric FC processes which can be observed 
in the future experiments. 

Concluding, we have shown that SUSY GUTs with the global horizontal 
symmetry $U(3)_H$ can predict very definite pattern of the SSB terms 
and naturally solve the flavor changing problem. 
(Unfortunatelly, the local $SU(3)_H$ could hardly serve for our purposes, 
since generally the large non-universal contributions would emerge 
from D-terms \cite{HKR}.)  
Note, however, that $U(3)_H$ can be reduced to its smaller subgroup 
without losing the key features of our approach. 
 In particular, one can consider 
the horizontal symmetry $G_1\times G_2 \times G_3$, 
with $G=U(1)$ or $Z_N$ independently acting on different families 
($i=1,2,3$), and assume invariance under the permutations 
$G_i\leftrightarrow G_j$. Needless to say that such 
models could have a better chance to be derived from the 
string theory. 

I would like to thank Antonio Masiero, Alex Pomarol and Zurab 
Tavartkiladze for useful discussions. 

%%%%%%%%%%%%%%%%%%%%%%%%%%%%%%%%%%%%%%%%%%%%%%%%%%%%%%%


\begin{thebibliography}{9}

\bibitem{PLB85} Z. Berezhiani, Phys. Lett. {\bf B150}, 177 (1985). 

\bibitem{ICTP} For a review, see  
Z. Berezhiani, hep-ph/9602325, to appear 
on Proc. ICTP Summer School (Trieste, July 1995). 

\bibitem{so10} G.W. Anderson et al., Phys. Rev. D {\bf 49}, 3660 (1994); 
K.S. Babu and R.N. Mohapatra, Phys. Rev. Lett. {\bf 74}, 2418 (1995); 
K.S. Babu and S.M. Barr, {\em ibid.} {\bf 75}, 2088 (1995); 
Z. Berezhiani, Phys. Lett. {\bf B355}, 178 (1995).

\bibitem{su6} Z. Berezhiani and G. Dvali,  Sov. Phys.  
-- Lebedev Inst. Reports {\bf 5}, 55 (1989); 
R. Ba\-rbieri et al., Nucl. Phys. {\bf B432}, 49 (1994); 
Z. Berezhiani, Phys. Lett. {\bf B355}, 481 (1995); 
Z. Berezhiani, C. Csaki, and L. Randall, Nucl. Phys. {\bf B444}, 61 (1995). 

\bibitem{BFS} R. Barbieri, S. Ferrara, and C. Savoy,
Phys. Lett. {\bf B119}, 343 (1982); 
P. Nath, R. Arnowitt, and A. Chamseddine, Phys. Rev. Lett. {\bf 49}, 
970 (1982); 
L. Hall, J. Lykken, and S. Weinberg, Phys. Rev. D {\bf 27}, 2359 (1983). 

\bibitem{HKR} 
L.J. Hall, V.A. Kostelecky, and S. Raby, Nucl. Phys. {\bf B267}, 415 (1986); 
R. Barbieri and L.J. Hall, Phys. Lett. {\bf B338}, 212 (1994). 

\bibitem{FN} 
C. Frogatt and H. Nielsen, Nucl. Phys. {\bf B147}, 277 (1979);  
Z. Berezhiani, Phys. Lett. {\bf B129}, 99 (1983); 
{\em ibid.} {\bf B150}, 177 (1985); 
S. Dimopoulos, {\em ibid.} {\bf B129}, 417 (1983). 

\bibitem{DP} 
S. Dimopoulos and A. Pomarol, Phys. Lett. {\bf B353}, 222 (1995);  
  Nucl. Phys. {\bf B453}, 83 (1995). 

\bibitem{NS} 
Y. Nir and N. Seiberg, Phys. Lett. {\bf B309}, 337 (1993); 
M. Leurer, Y. Nir, and N. Seiberg, Nucl. Phys. {\bf B398}, 319 (1993); 
E. Dudas, S. Pokorski, and C. Savoy, Phys. Lett. {\bf B369}, 255 (1995);  
%\bibitem{PT} 
M. Dine, A. Kagan, and R. Leigh, Phys. Rev. D {\bf 48}, 4269 (1993); 
A. Pomarol and D. Tommasini, Nucl. Phys. {\bf B466}, 3 (1996); 
R. Barbieri, G. Dvali, and L.J. Hall, LBL-38065 (1996); 
R. Barbieri and L.J. Hall, LBL-38381 (1996).  

\bibitem{su3H} J. Chkareuli, JETP Lett. {\bf 32}, 684 (1980); 
Z. Berezhiani and J. Chkareuli, JETP Lett. {\bf 35}, 612 (1982); 
Yad. Fiz. {\bf 37}, 1043 (1983); 
Z. Berezhiani and M. Khlopov, Yad. Fiz. {\bf 51}, 1157,1479 (1990); 
Z. Phys. C {\bf 49}, 73 (1991). 

\bibitem{Fr} H. Fritzsch, Nucl. Phys. {\bf B155}, 189 (1979). 

\bibitem{susy96} Although the exact Fritzsch texture for fermion 
masses is already excluded by experiment, for its viable modification 
see Z. Berezhiani, hep-ph/9607363. 

\bibitem{Rattazzi} 
Strictly speaking, the seesaw approximation may not be good for the 
third family, since $\la_t\sim 1$ and thus $C\sim M$. 
Then one has to use more precise formulas -- see e.g. in 
Z. Berezhiani and R. Rattazzi, Nucl. Phys. {\bf B407}, 249 (1993). 
E.g., for $\la_t\simeq 1$ and say $g'\simeq 2$, 
constants $\la_{t,b,\tau}$ will get few percent corrections.  
The strong deviations from the seesaw limit will be discussed elsewhere. 

\bibitem{HOP} 
It would be more economic to take all horizontal scalars $\chi_n$ 
as $SU(5)$ singlets, and to assume e.g. that 
$\chi_3$ has a direct coupling $\chi_3 F_c f_{10}$ 
while $\chi_2$ couples via the higher order operator 
$\frac{\Phi}{\La}\chi_2 F_c f_{10}$, 
where $\Phi$ is a Higgs 24-plet of $SU(5)$.  
In general, this would violate the soft mass universality 
at the scale $M$, but not necessarily.  
For example, the first operator can be effectively induced by 
integrating out the additional fermion states 
$\Psi(10,3) + \Psi_c(\ov{10},\bar3)$  from the superpotential  
$\chi_2 F_c \Psi + \Lambda \Psi \Psi_c + \Phi \Psi_c f_{10}$, 
with $\La\gg M,M_X$.  As far as mixing between the states 
$f_{10}$ and $\Psi$ is universal in families, at least in the leading 
approximation, no additional FC effects will be induced.  

\bibitem{prop} 
Alternatively, one could assume that $A$-terms in (\ref{tri-F}) 
are all vanishing due to the SUSY breaking specifics.  
It can be also envisaged that all fields $\chi_n$ actually emerge 
from a single supermultiplet of some underlying theory, so that 
$a'_n\propto a_n$.  Then the matrices 
$\hat{\rho}_f$ and $\hat{\kappa}_f$ are exactly proportional  
and thus $\delta_A$ can be absorbed into $\delta$. 

\bibitem{Turok} 
M. Joyce and N. Turok, Nucl. Phys. {\bf B416}, 389 (1994). 

\bibitem{Kim} For review, see J.E. Kim, Phys. Rep. {\bf 150}, 1 (1987). 


\end{thebibliography}
\end{document}